\documentclass[aps,preprintnumbers,eqsecnum,amsmath,amssymb,showpacs,nofootinbib]{revtex4}
\usepackage{graphicx}
\usepackage{dcolumn}
\usepackage{bm}
\usepackage{epsfig}

\begin{document} 

\title{\boldmath Accurate decay-constant ratios $f_{B^*}/f_B$ and $f_{B_s^*}/f_{B_s}$ from Borel QCD sum rules}
\author{Wolfgang Lucha$^a$, Dmitri Melikhov$^{a,b,c}$, and Silvano Simula$^d$}
\affiliation{$^a$Institute for High Energy Physics, Austrian
Academy of Sciences, Nikolsdorfergasse 18, A-1050 Vienna,
Austria\\ $^b$D.~V.~Skobeltsyn Institute of Nuclear Physics,
M.~V.~Lomonosov Moscow State University, 119991, Moscow, Russia\\
$^c$Faculty of Physics, University of Vienna, Boltzmanngasse 5,
A-1090 Vienna, Austria\\ $^d$INFN, Sezione di Roma III, Via della
Vasca Navale 84, I-00146 Roma, Italy}\date{\today}

\begin{abstract}We present our analysis of the decay constants of
the beauty vector mesons $B^*$ and $B^*_s$ within the framework of
dispersive sum rules for the two-point correlator of vector
currents in QCD. While the decay constants of the vector mesons
$f_{B^*}$ and $f_{B_s^*}$---similar to the decay constants of the
pseudoscalar mesons $f_B$ and $f_{B_s}$---individually have large
uncertainties induced by theory parameters not known with a
satisfactory precision, these uncertainties almost entirely cancel
out in the ratios of vector over pseudoscalar~decay constants.
These ratios may be thus predicted with very high accuracy due to
the good control over the systematic uncertainties of the decay
constants gained upon application of our hadron-parameter
extraction algorithm. Our final results read
$f_{B^*}/f_B=0.944\pm0.011_{\rm OPE}\pm0.018_{\rm syst}$ and
$f_{B_s^*}/f_{B_s}=0.947\pm0.023_{\rm OPE}\pm0.020_{\rm syst}$.
Thus, both $f_{B^*}/f_B$ and $f_{B_s^*}/f_{B_s}$ are less than
unity~at 2.5$\sigma$ and 2$\sigma$ level,
respectively.\end{abstract}

\pacs{11.55.Hx, 12.38.Lg, 03.65.Ge} \maketitle

\section{Introduction}The QCD sum-rule approach
\cite{svz,aliev,rubinstein}, based on the application of Wilson's
operator product expansion (OPE) to the properties of individual
hadrons, has been extensively used for predicting heavy-meson
decay constants. An~important finding of these analyses was the
strong sensitivity of the decay constants to the values of the
input OPE parameters and to the prescription of fixing the
effective continuum threshold \cite{lms_1}. The latter governs the
accuracy of the quark--hadron duality approximation and, to a
large extent, determines the extracted value of the decay
constant. Even~if~the parameters of the truncated OPE are known
with arbitrarily high precision, the decay constants may be
predicted with only limited accuracy, which we refer to as their
systematic uncertainty. In a series of papers \cite{lms_new}, we
have~formulated a new algorithm for fixing the effective threshold
within Borel QCD sum rules and for obtaining reliable
estimates~for the systematic uncertainties. This procedure opened
the possibility to provide predictions for the decay
constants~with a controlled accuracy \cite{lms_fp,lms_fD}.

Here, we study the decay constants of the vector beauty mesons
$f_{B^*}$ and $f_{B^*_s}$. As is already known from the analysis
of the decay constants of the pseudoscalar mesons $B$ and $B_s$
\cite{lms_fp}, the OPE uncertainties in the obtained
predictions~are rather large. The same occurs also for the $B^*$
and $B_s^*$ mesons. However, the OPE uncertainties to a great
extent cancel out in the ratios of the decay constants of vector
and pseudoscalar beauty mesons. An important result reported here
is that the systematic uncertainties of the decay constants are
rather small and well under control. Therefore, these ratios are
predicted with a very good accuracy. It should be taken into
account that we address a rather subtle effect at a few-percent
level; a priori, it is not clear whether QCD sum rules are, in
principle, capable to provide~theoretical predictions at this
level of accuracy. Obviously, the control over the systematics is
becoming crucial.

The ratio of the decay constants of vector over pseudoscalar heavy
mesons is an interesting quantity: it is known to be unity in the
heavy-quark limit and to approach this limit from below because of
the radiative corrections \cite{neubert}. For beauty mesons, the
few existing sum-rule analyses (which, however, could not gain
good control over the systematic uncertainties) reported
$f_{B^*}/f_B$ slightly above unity \cite{narison2014,kh}.
Constituent-quark models typically also yield $f_{B^*}/f_B>1$
\cite{faustov}. A similar conclusion has been reached by
interpolation of the lattice data from the charm-quark mass region
to the beauty-quark mass \cite{becirevic2014}.

The first indication that this ratio for beauty mesons is below
unity was given in our papers \cite{lms2014}. Recently, HPQCD
\cite{hpqcd2015} also reported an accurate value of
$f_{B^*}/f_B<1$, in excellent agreement with the results of
\cite{lms2014}. The analysis of \cite{lms2014}, although
conclusively indicating $f_{B^*}/f_B<1$, observed an unpleasant
dependence of the extracted decay constants of the vector beauty
mesons on the renormalization scale $\mu$ chosen for the
evaluation of the vector correlation function. This analysis
solves the problem of the sensitivity to the choice of the scale
$\mu$ by improving the extraction procedures for the decay
constants and arrives at new predictions stable with respect to
the choice of $\mu$. Our detailed results read
\begin{eqnarray}
f_{B^*}/f_B=0.944\pm0.011_{\rm OPE}\pm0.018_{\rm syst},\qquad
f_{B_s^*}/f_{B_s}=0.947\pm0.023_{\rm OPE}\pm0.020_{\rm syst},
\end{eqnarray}
in more than excellent agreement with the latest results from
lattice QCD \cite{hpqcd2015}. Let us emphasize once more that~the
OPE uncertainties cancel to a large extent in the above ratios.
Thus, decisive for obtaining an accurate sum-rule~result is our
capability to control the systematic uncertainties of the QCD
sum-rule method.

\section{QCD vector correlator and sum rule for vector-meson decay
constant $f_V$}The decay constants of ground-state vector mesons
may be extracted by analyzing the two-point correlation function
\begin{eqnarray}
\label{1.1}i\int d^4x\,e^{ipx}
\langle0|T\!\left(j_\mu(x)j^\dagger_\nu(0)\right)|0\rangle=
\left(-g_{\mu\nu}+\frac{p_\mu p_\nu}{p^2}\right)\Pi(p^2)
+\frac{p_\mu p_\nu}{p^2}\Pi_L(p^2)
\end{eqnarray}
of the heavy--light vector currents for a heavy quark $Q$ of mass
$m_Q$ and a light quark $q$ of mass $m$,
\begin{eqnarray}
j_\mu(x)=\bar q(x)\gamma_\mu Q(x),
\end{eqnarray}
or, more precisely, the Borel transform of its transverse
structure $\Pi(p^2)$ to the Borel variable $\tau$, $\Pi(\tau)$.
Equating~$\Pi(\tau)$~as calculated within QCD and the expression
obtained by insertion of a complete set of hadron states yields
the sum~rule
\begin{eqnarray}
\label{pitau}\Pi(\tau)=f^2_{V}M_V^2e^{-M_V^2\tau}
+\int\limits_{s_{\rm phys}}^{\infty}ds\,e^{-s\tau}\rho_{\rm
hadr}(s)=\int\limits^\infty_{(m_Q+m)^2}ds\,e^{-s\tau}\rho_{\rm
pert}(s,\mu)+\Pi_{\rm power}(\tau,\mu).
\end{eqnarray}
Here, $M_V$ labels the mass, $f_V$ the decay constant, and
$\varepsilon_\mu(p)$ the polarization vector of the vector meson
$V$ under~study:
\begin{eqnarray}
\label{decay_constant}\langle0|\bar q\gamma_\mu Q|V(p)\rangle
=f_VM_{V}\varepsilon_\mu(p).
\end{eqnarray}
For the correlator (\ref{1.1}), $s_{\rm phys}=(M_{P}+M_\pi)^2$ is
the physical continuum threshold, wherein $M_{P}$ denotes the mass
of~the lightest pseudoscalar meson containing $Q$. For large
values of $\tau$, the ground state dominates the
correlator~and~thus its properties may be extracted from the
correlation function (\ref{1.1}).

In perturbation theory, the correlation function is found as an
expansion in powers of the strong coupling ``constant''
$\alpha_{\rm s}(\mu)$. The best known three-loop perturbative
spectral density has been calculated in \cite{chetyrkin} in terms
of the pole mass~of the heavy quark $Q$ (that is, in the present
case, $M_b$) and for a massless second quark [hereafter, we use
the abbreviation $a(\nu)=\alpha_{\rm s}(\nu)/\pi$, where
$\alpha_{\rm s}(\nu)$ is the running coupling at renormalization
scale $\nu$ in the $\overline{\rm MS}$ scheme]:
\begin{eqnarray}
\label{rhopert}\rho_{\rm pert}(s)=\rho^{(0)}(s,M_b)
+a(\nu)\rho^{(1)}(s,M_b)+a^2(\nu)\rho^{(2)}(s,M_b,\mu)+\cdots.
\end{eqnarray}
For both quarks having nonzero masses, the two-loop spectral
density in terms of their pole masses was
obtained~in~\cite{rubinstein}.

The power corrections are also separately scale-independent; their
explicit expressions can be found in \cite{jamin,kh}. For
instance, for pseudoscalar ($P$) and vector ($V$) currents the
quark-condensate contributions may be written in the~form
\begin{eqnarray}
\label{pipower}\Pi^P_{\rm power}(\tau)
&=&-\overline{m}_b(\nu)\langle\bar qq(\nu)\rangle M_b^2
\left[\exp(-M_b^2\tau)\left(1+\frac{3}{2}C_F a\right)-\frac{3}{2}
C_Fa\,\Gamma(0,M_b^2\tau)\right],\\ \Pi^V_{\rm power}(\tau)
&=&-\overline{m}_b(\nu)\langle\bar qq(\nu)\rangle
\left[\exp(-M_b^2\tau)\left(1+\frac{1}{2}C_F a\right)
+\frac{1}{2}C_F a\, M_b^2\tau \,\Gamma(-1,M_b^2\tau)\right],
\end{eqnarray}
where $\overline{m}_b(\nu)$ is the $b$-quark $\overline{\rm MS}$
mass at renormalization scale $\nu$,
$\overline{m}_b(\nu)\langle\bar qq(\nu)\rangle$ is a
scale-independent combination,~and $\Gamma(n,z)$ is the incomplete
gamma function \cite{AS}.

However, even if the lowest-order contributions to the
perturbative expansion and the vacuum condensates of lowest
dimensions are known to good accuracy, a truncated OPE does not
allow one to calculate the correlator for sufficiently large
$\tau$, such that the continuum states give a negligible
contribution to $\Pi(\tau)$ in the corresponding range of $\tau$.
In~order to get rid of the continuum contribution, the concept of
duality is invoked: Perturbative-QCD spectral density $\rho_{\rm
pert}(s)$ and hadron spectral density $\rho_{\rm hadr}(s)$
resemble each other at large values of $s$; thus, for values of
the integration lower limit $\bar s$ chosen sufficiently large,
that is to say, (far) above the resonance region, one arrives at
the duality relation
\begin{eqnarray}
\label{duality1} \int\limits_{\bar s}^{\infty}ds\,e^{-s
\tau}\rho_{\rm hadr}(s)=\int\limits_{\bar s}^{\infty}ds\,e^{-s
\tau}\rho_{\rm pert}(s).
\end{eqnarray}
Now, in order to express the hadron continuum contribution in
terms of the perturbative contribution, the relation
(\ref{duality1}) should be extended down to the hadronic or
physical threshold $s_{\rm phys}$. However, since the spectral
densities~$\rho_{\rm pert}(s)$~and $\rho_{\rm hadr}(s)$ obviously
differ in the region near $s_{\rm phys}$, one can reasonably only
expect to obtain a relationship of the form
\begin{eqnarray}
\label{duality}
\int\limits_{s_{\rm phys}}^{\infty} ds\, e^{-s
\tau} \rho_{\rm hadr}(s) = \int\limits_{s_{\rm
eff}(\tau)}^{\infty} ds\, e^{-s \tau} \rho_{\rm pert}(s),
\end{eqnarray}
where the effective threshold $s_{\rm eff}(\tau)$ is clearly
different from the physical threshold $s_{\rm phys}$, $s_{\rm
eff}(\tau)\ne s_{\rm phys}$, and, moreover, must be a function of
the Borel parameter $\tau$ \cite{lms_1,lms_new}. By virtue of
(\ref{duality}), we may hence rewrite the QCD sum rule
(\ref{pitau})~as
\begin{eqnarray}
\label{sr} f_V^2 M_V^2 e^{-M_V^2\tau}= \int\limits^{s_{\rm
eff}(\tau)}_{(m_Q+m)^2} ds\, e^{-s\tau}\rho_{\rm pert}(s,\mu) +
\Pi_{\rm power}(\tau,\mu) \equiv \Pi_{\rm dual}(\tau,s_{\rm eff}(\tau)).
\end{eqnarray}
We refer to the right-hand side of this relation as the {\it dual
correlator\/}, and to the masses and decay constants extracted
from this expression as the corresponding \emph{dual\/}
quantities. In addition to $\rho_{\rm pert}(s,\mu)$ and $\Pi_{\rm
power}(\tau,\mu)$, the extraction~of $f_V$ requires, as further
input, a criterion that fixes the functional behaviour of the
effective continuum threshold~$s_{\rm eff}(\tau)$.

We shall demonstrate that QCD sum rules allow a very satisfactory
extraction of the vector-meson decay constants, with an accuracy
that is certainly competitive to that found within the framework
of lattice QCD.

\section{OPE and choice of renormalization scheme and scale for
heavy-quark mass}The starting point of our discussion is the OPE
for the correlator (\ref{1.1}). The three-loop perturbative
spectral density $\rho_{\rm pert}(s,M)$ was calculated in
\cite{chetyrkin} in terms of the pole mass~of the heavy quark. A
nice feature of the pole-mass OPE is that each of the known
perturbative contributions to the dual correlator is positive.
Unfortunately, the pole-mass~OPE does not provide a visible
hierarchy of the perturbative contributions to the extracted
predictions, which raises doubts whether the
$O(\alpha_s^2)$-truncated pole-mass OPE is indeed a good starting
point for a reliable analysis of decay constants.

A well-known remedy is to reorganize the perturbative expansion in
terms of the $b$-quark running $\overline{\rm MS}$ mass
$\overline{m}_b(\mu)$, related (in the notations of \cite{jamin})
to the corresponding pole mass $M_b$ by
\begin{eqnarray}
\label{a2} M_b=
\overline{m}_b(\mu)/\left(1+a(\mu)r^{(1)}_m+a^2(\mu)r_m^{(2)}\right)
+O(a^3).
\end{eqnarray}
The spectral densities in the $\overline{\rm MS}$ scheme are found
by expanding the pole-mass spectral densities in powers
of~$a(\mu)$ and omitting terms of order $O(a^3)$ and higher;
starting at order $O(a)$, they contain two parts: the ``genuine''
part~from \cite{chetyrkin} and the part induced by the lower
perturbative orders when expanding the pole mass in terms of the
running~mass. By this, however, due to the truncation of the
perturbative series, one gets an explicit (unphysical) dependence
of~the dual correlator and of the extracted decay constant on the
scale $\mu$. In principle, any scale should be equivalently~good.
In practice, however, the distinctness of the hierarchy of the
perturbative contributions to the dual correlator depends on the
precise choice of the scale. This opens a possibility of choosing
the scale $\mu$ such that the hierarchy of the new perturbative
expansion is improved.

\begin{figure}[!t]
\begin{tabular}{cc}
\includegraphics[width=7.5581cm]{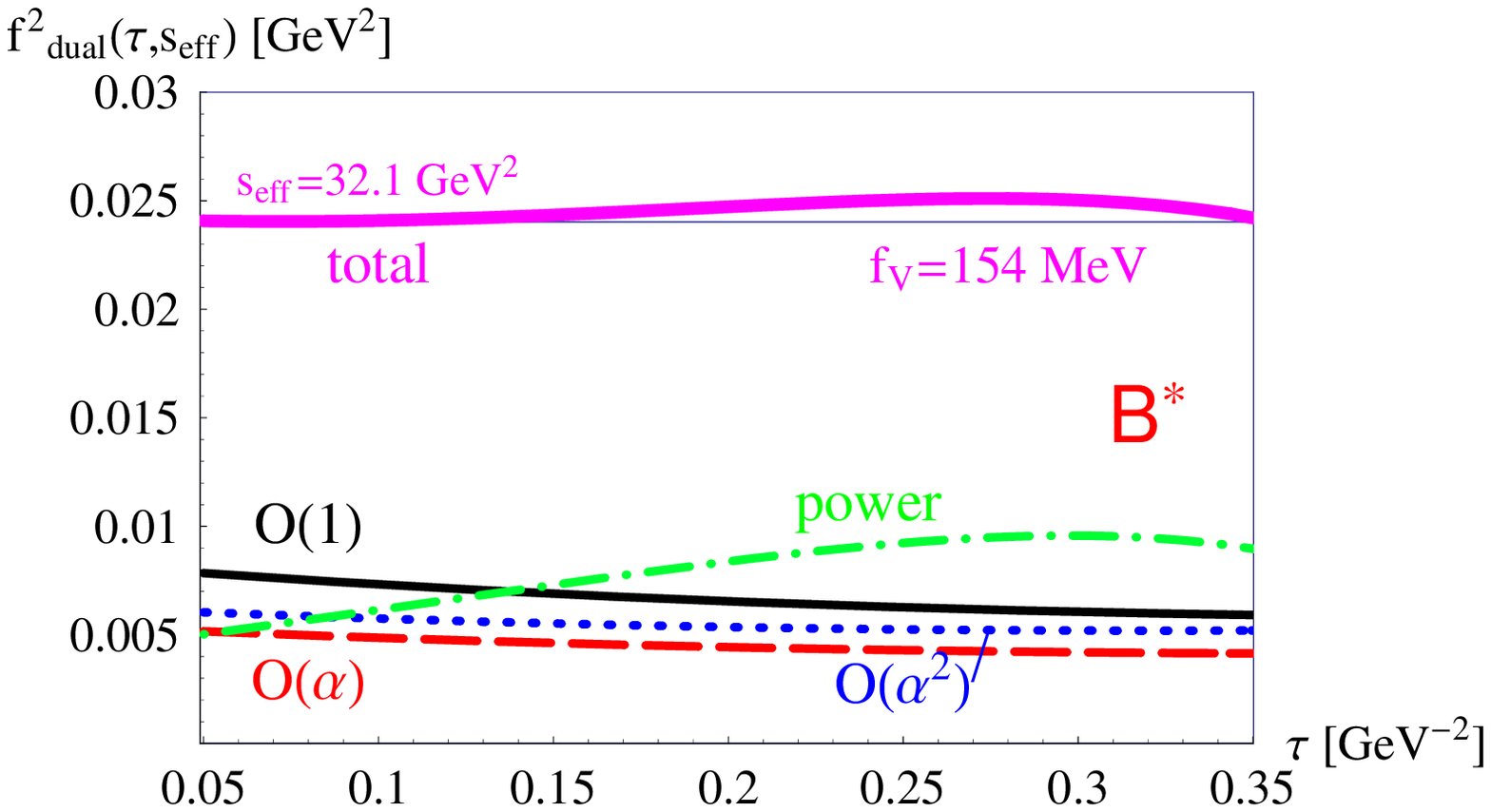} &
\includegraphics[width=7.5581cm]{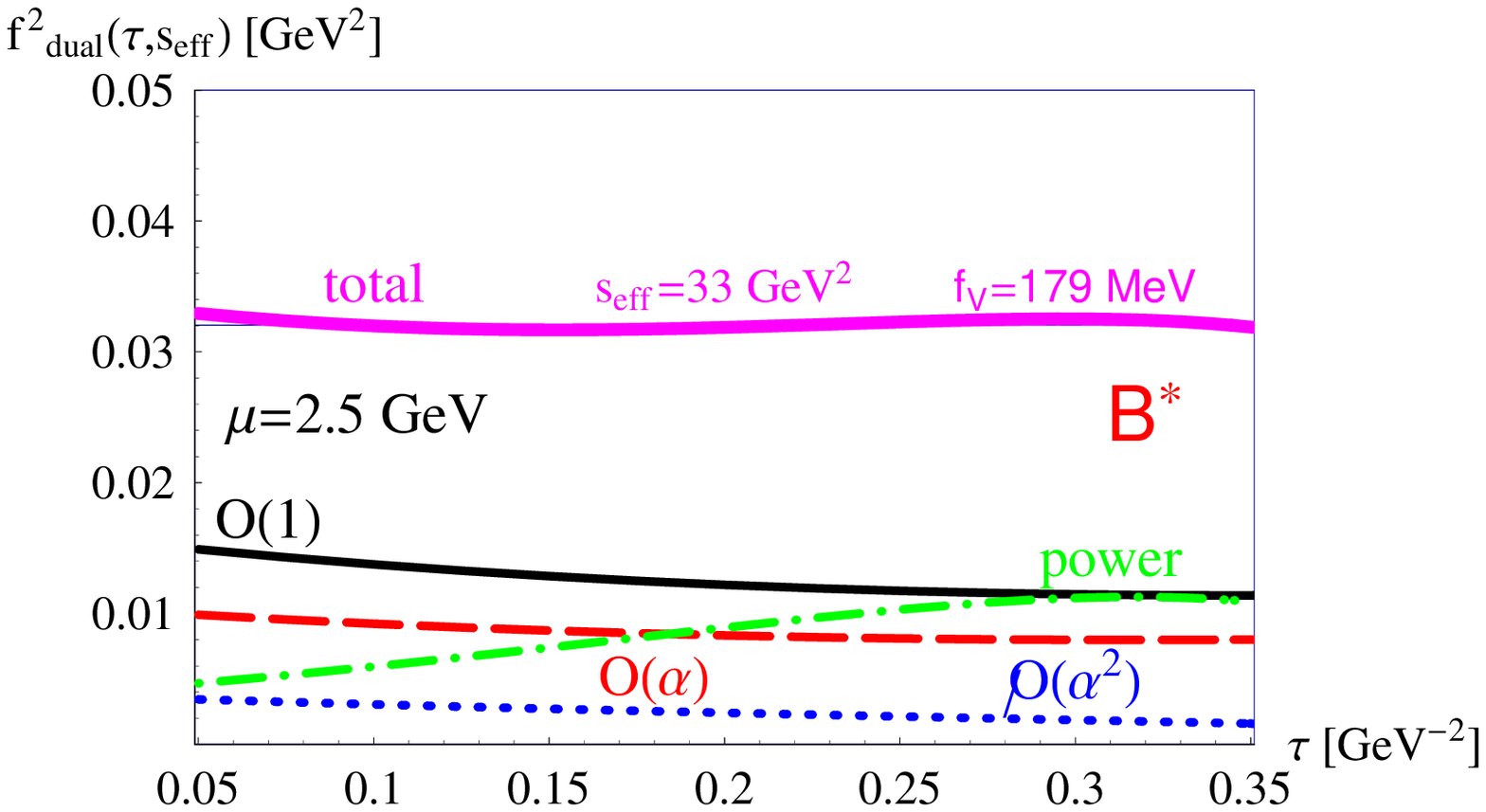} \\
\includegraphics[width=7.5581cm]{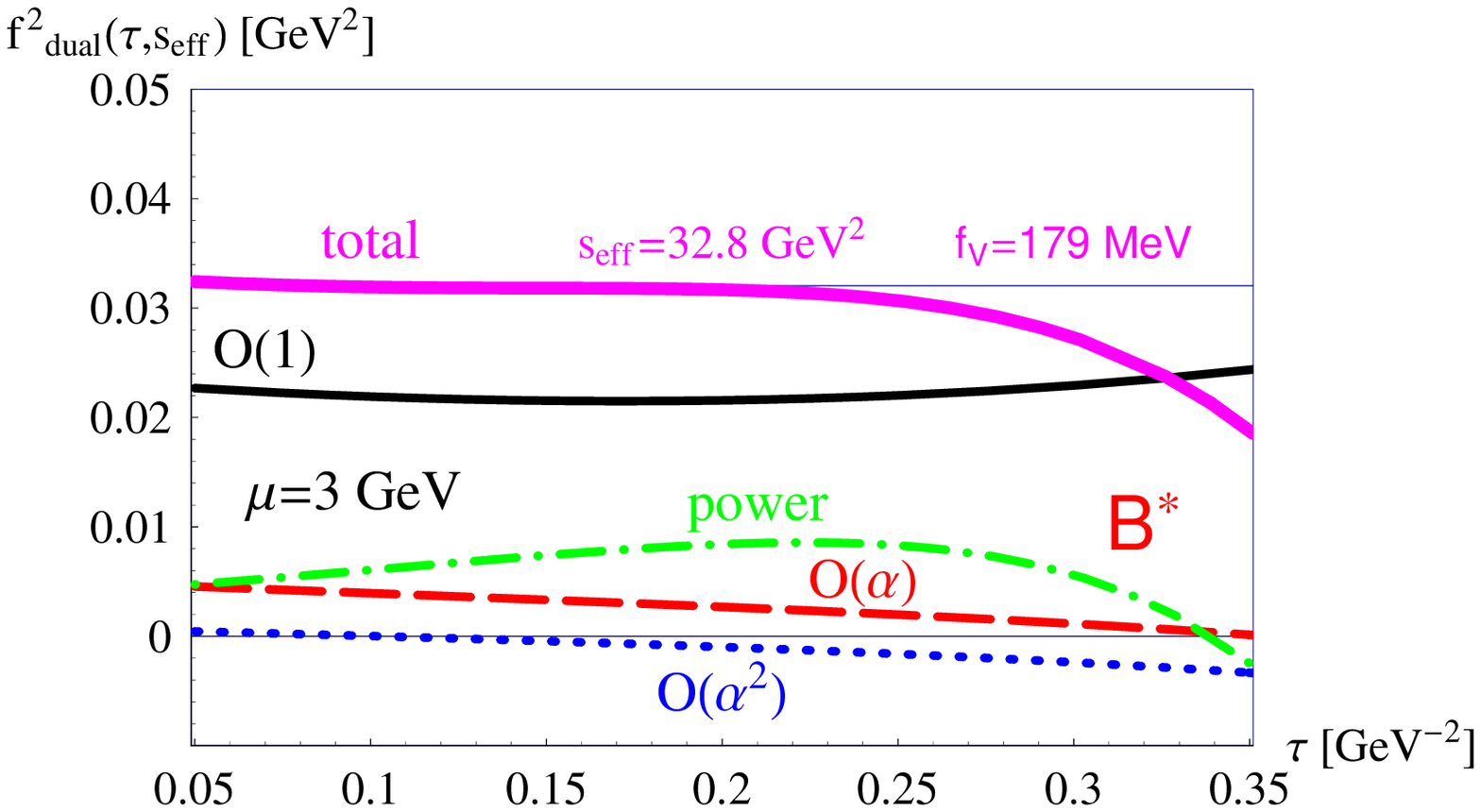} &
\includegraphics[width=7.5581cm]{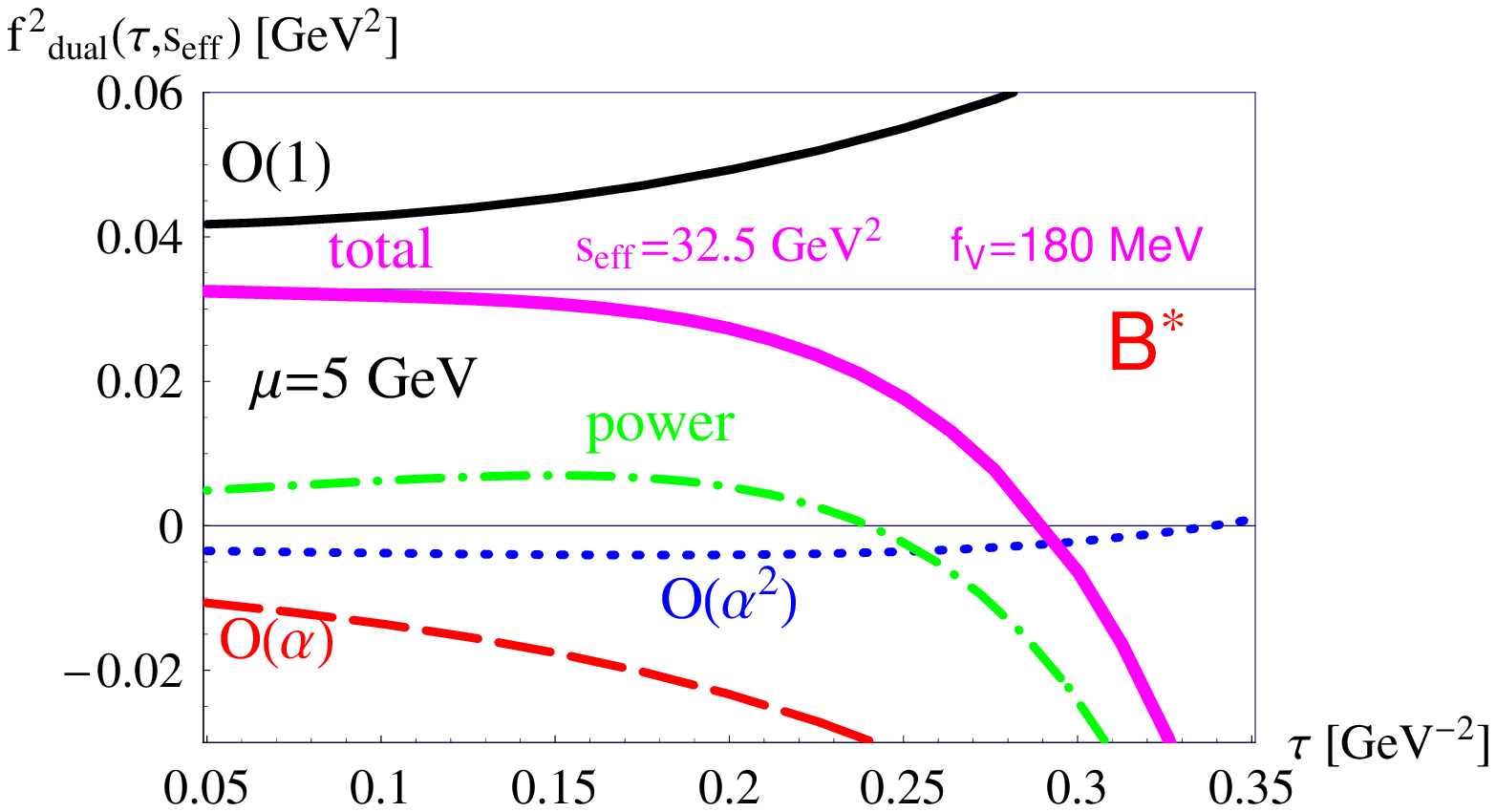}
\end{tabular}
\caption{\label{Plot:1a}QCD sum-rule estimates of the $B^*$-meson
decay constant using the pole-mass OPE (a) and the running-mass
OPE at the renormalization scales $\mu=2.5$ GeV (b), $\mu=3$ GeV
(c), and $\mu=5$ GeV (d). The running-mass OPE for
$\overline{m}_b(\overline{m}_b)=4.247$~GeV is shown. The pole-mass
OPE employs the corresponding two-loop pole mass $M_b=4.87$ GeV.
For each case, separately, a constant effective continuum
threshold~$s_{\rm eff}$ is determined by requiring maximal
stability of the predicted decay constant in a Borel window of the
maximal width $0.05\le\tau\ ({\rm GeV}^{-2})\le0.15$. Bold lines
(lilac)---total findings, solid lines
(black)---$O(1)$~contributions; dashed lines
(red)---$O(\alpha_{\rm s})$ contributions; dotted lines
(blue)---$O(\alpha_{\rm s}^2)$ contributions; dot-dashed lines
(green)---power contributions.}
\end{figure}

\begin{figure}[!ht]
\begin{tabular}{cc}
\includegraphics[width=7.5581cm]{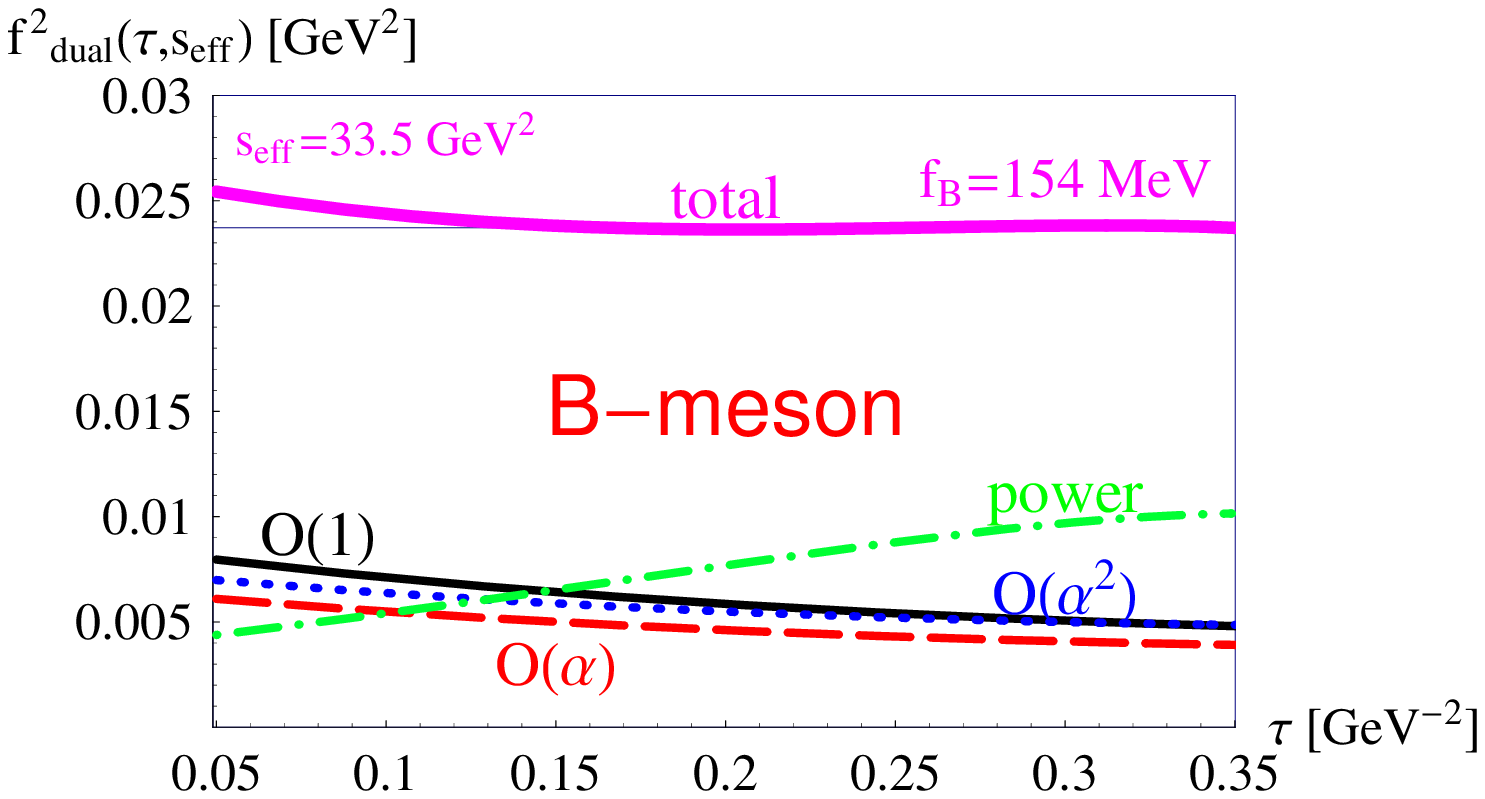} &
\includegraphics[width=7.5581cm]{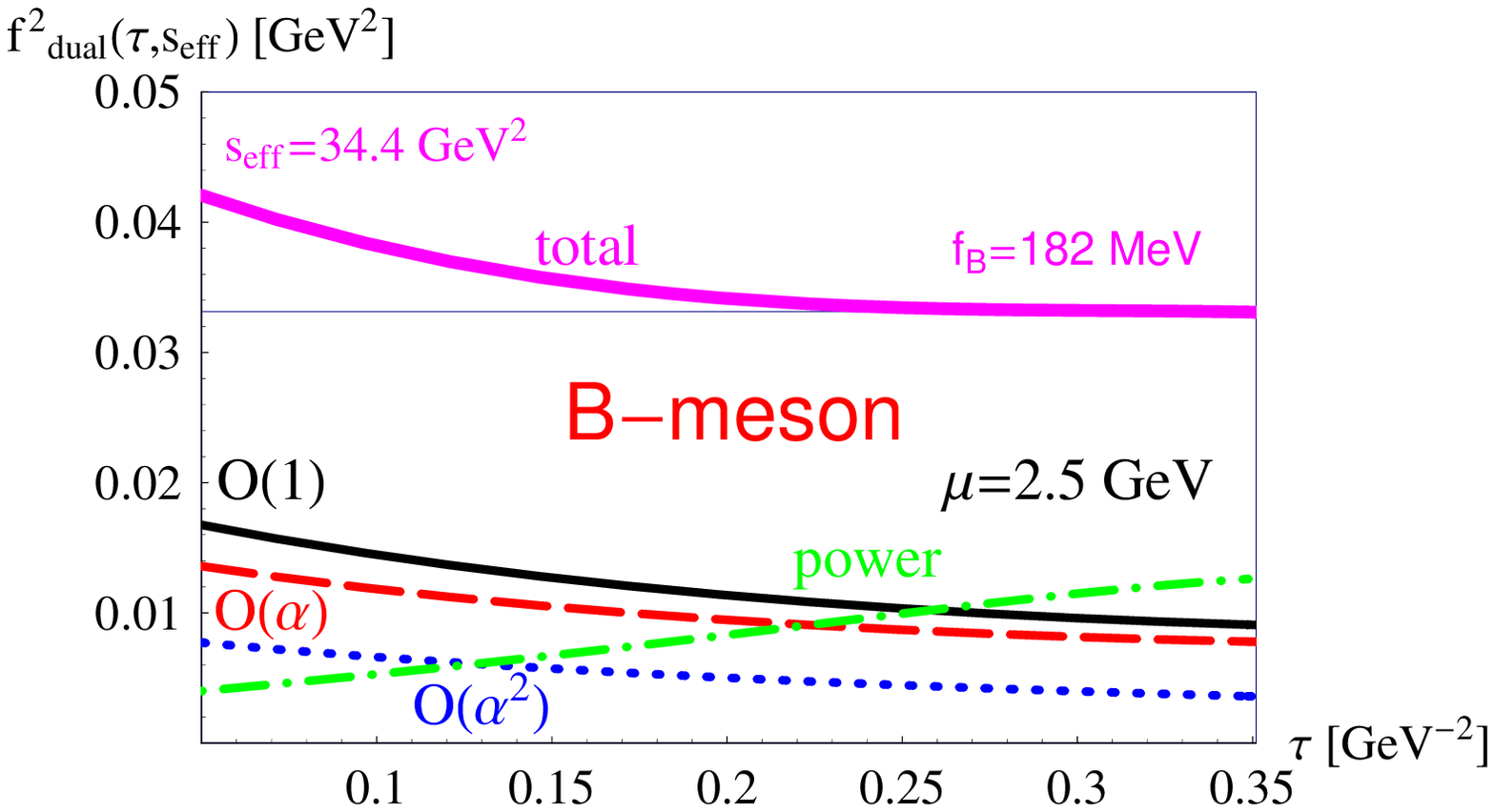} \\
\includegraphics[width=7.5581cm]{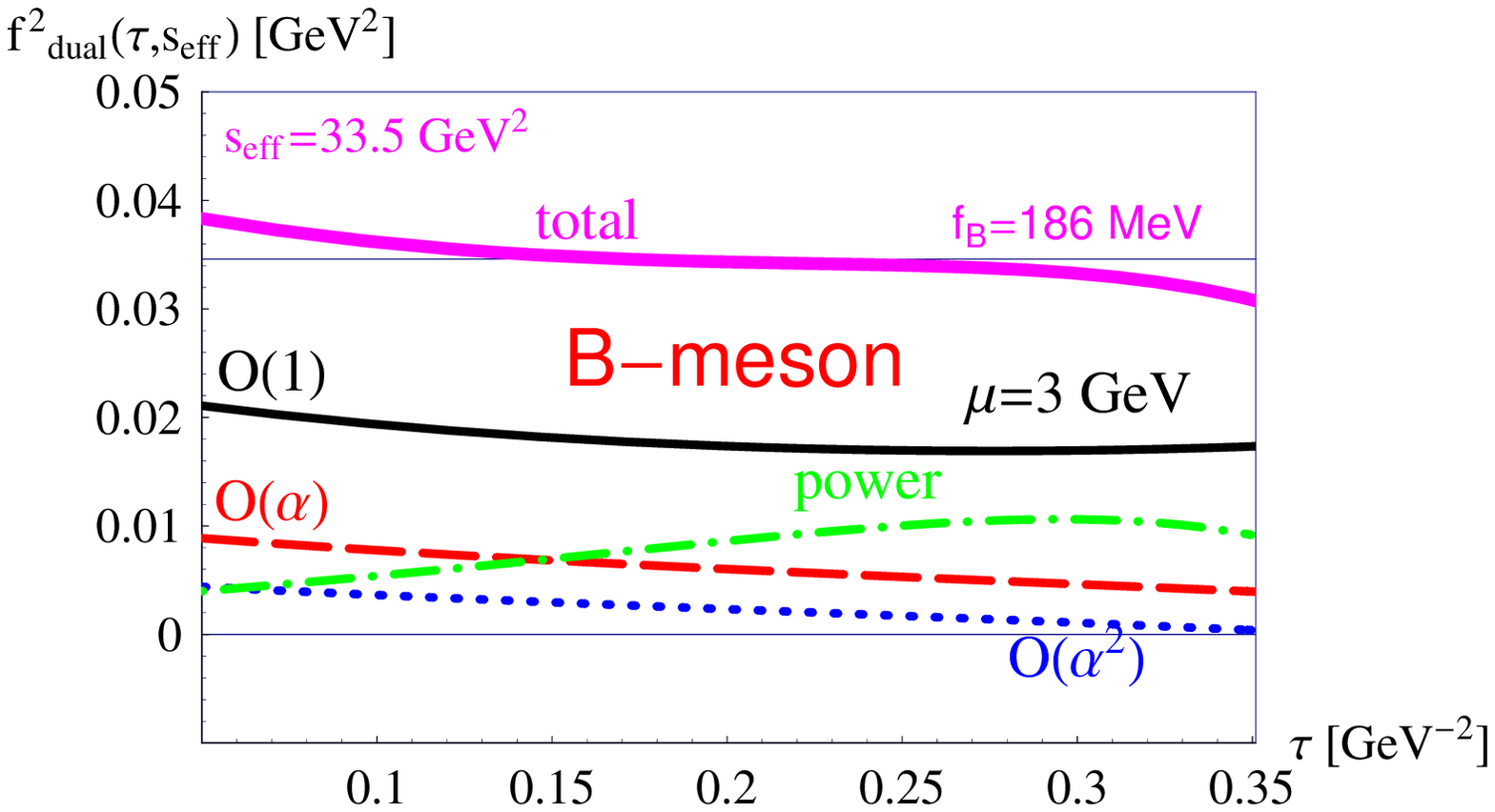} &
\includegraphics[width=7.5581cm]{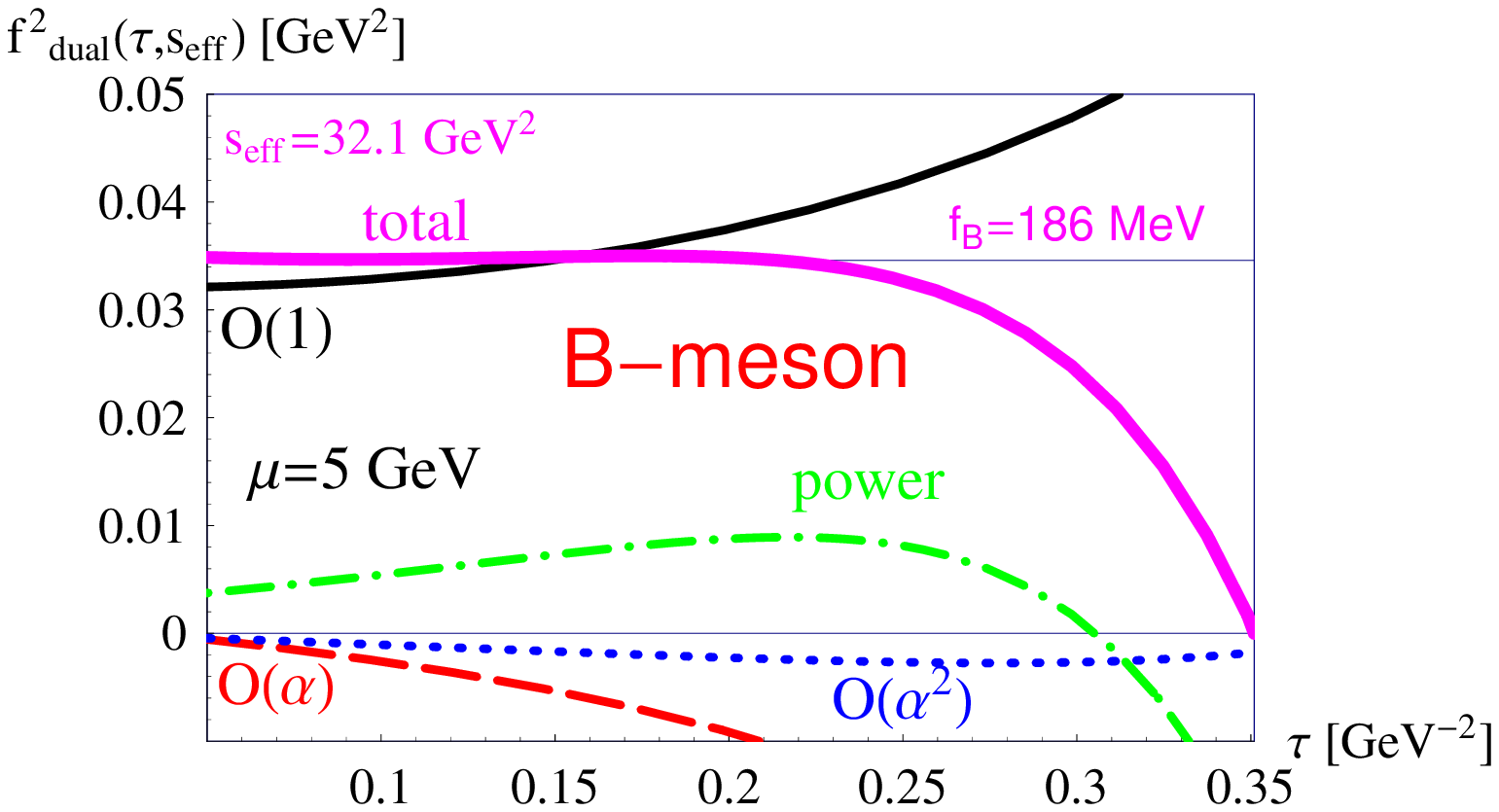}
\end{tabular}
\caption{\label{Plot:1b} Same as Figure \ref{Plot:1a} but for the
$B$ meson.}
\end{figure}

Figures \ref{Plot:1a} and \ref{Plot:1b} depict the
dual decay constants of the $B^*$ and $B$ mesons, respectively. 
For the~$b$-quark $\overline{\rm MS}$ mass, we use the value
determined in \cite{ourmb} by matching our QCD sum-rule results
for $f_B$ to those of lattice~QCD:\footnote{As shown in
\cite{ourmb}, the PGD average
$\overline{m}_b(\overline{m}_b)=(4.18\pm0.030)$ GeV \cite{pdg}
(see also \cite{erler} for a recent overview of the $b$-quark mass
results)~leads to a considerably larger value of $f_B$,
incompatible with the latest lattice-QCD results. However, the
precise value of $\overline{m}_b(\overline{m}_b)$ has~negligible
impact on the ratio of the decay constants of vector and
pseudoscalar mesons.}
\begin{eqnarray}
\label{ourmb}\overline{m}_b(\overline{m}_b)=(4.247\pm0.034)\ {\rm GeV}.
\end{eqnarray}
The numerical values adopted for other relevant OPE parameters are
\cite{jamin,ourmb,lms_fD,ms2014}
\begin{eqnarray}
\label{Table:1} 
&& m(2\;{\rm GeV})=(3.42\pm0.09)\;{\rm MeV}, \quad m_s(2\;{\rm GeV})=(93.8\pm2.4)\;{\rm MeV},\quad \alpha_{\rm s}(M_Z)=0.1184\pm0.0020, \nonumber\\
&&\left\langle\frac{\alpha_{\rm s}}{\pi}GG\right\rangle=(0.024\pm0.012)\;{\rm GeV}^4,\quad 
\langle\bar qq\rangle(2\;{\rm GeV})=-[(267\pm17)\;{\rm MeV}]^3,\quad \frac{\langle\bar ss\rangle(2\;{\rm
GeV})}{\langle\bar qq\rangle(2\;{\rm GeV})}=0.8\pm0.3.
\end{eqnarray}
The purpose of Figs.~\ref{Plot:1a} and \ref{Plot:1b} is the illustration of the main features of the dual correlators
(\ref{sr}), therefore the QCD sum-rule estimates shown here are obtained for 
a $\tau$-independent effective threshold: $s_{\rm eff}=\mbox{const}$. 
The numerical value of the latter is, in each
case, found by requiring maximal stability of the extracted decay
constant in the Borel~window. We emphasize that our results for the decay constants reported in the next 
Sections are obtained using the $\tau$-dependent effective thresholds. 

From Figs.~\ref{Plot:1a} and \ref{Plot:1b}, we conclude that the
$O(\alpha_s^2)$-truncated pole-mass OPE exhibits no hierarchy of
the perturbative expansion and better should not be used.
Unfortunately, the hierarchy of the running-mass OPE is also not
guaranteed automatically and depends strongly on the scale $\mu$.

Let us define a scale $\hat\mu$ by demanding
$M_b=\overline{m}_b(\hat\mu)$. From the $O(a^2)$ relation between
$\overline{\rm MS}$ and pole mass,~we find $\hat\mu\approx2.23$
GeV. At this scale, the perturbative hierarchy of the
$\overline{\rm MS}$ expansion is worse than that of the pole-mass
expansion because the $O(1)$ spectral densities coincide, whereas
the $O(\alpha_s)$ spectral density in the $\overline{\rm MS}$
scheme receives a positive contribution compared to the pole-mass
scheme. For lower scales $\mu<\hat\mu$, the hierarchy of the
$\overline{\rm MS}$ expansion gets worse with decreasing $\mu$.
For higher scales $\mu>\hat\mu$, first the hierarchy of the
$\overline{\rm MS}$-expansion improves with rising $\mu$
(Figs.~\ref{Plot:1a} and \ref{Plot:1b}). However, as the scale
$\mu$ becomes sufficiently larger than $\hat\mu$, the ``induced''
contributions, which mainly reflect the bad-behaved expansion of
the pole mass in terms of the running mass, start to dominate over
the ``genuine'' contributions. This is evident in
Figs.~\ref{Plot:1a} and \ref{Plot:1b}: at $\mu=5$ GeV, the $O(1)$
contribution to the dual correlator rises~steeply with $\tau$,
whereas the $O(a)$ contribution becomes negative in order to
compensate the rising $O(1)$ contribution. Finally, for large
values of $\mu$ we mainly observe a compensation between the
``induced'' contributions. We may expect in this case the accuracy
of the expansion to deteriorate.

Figures \ref{Plot:1a} and \ref{Plot:1b} also reveal an essential
difference between pseudoscalar and vector correlators: at the
same scale $\mu$, the good reproduction of the observed mass of
the vector meson requires lower values of $\tau$ compared to its
pseudoscalar partner. This implies that the Borel window for the
vector correlator should be chosen at lower values of $\tau$ than
the corresponding window for the pseudoscalar correlator.
Moreover, for $\mu\gtrsim5$--6 GeV the vector-meson mass cannot be
reproduced in a reasonably broad $\tau$ window and so the QCD sum
rule cannot predict the vector-meson decay~constant.

For the present analysis, we thus choose as range of scales
$\mu=3$--5 GeV: On the one hand, in this range we observe a
reasonable hierarchy of the perturbative contributions to the
correlator. On the other hand, we shall see that for~this range of
scales one can find sufficiently broad $\tau$ windows where the
decay constants may be reliably extracted by our algorithm. For
the vector mesons, the upper bound of this window depends on
$\mu$.

\section{Extraction of the beauty-meson decay constants from our
QCD sum rule}In order to extract the decay constants, we first
have to find a $\tau$ window such that the OPE provides a
sufficiently accurate description~of the exact correlator (i.e.,
all higher-order radiative and power corrections are under
control). Next, we must determine the $\tau$ dependence of the
effective threshold $s_{\rm eff}(\tau)$. The appropriate algorithm
was developed and verified within quantum-mechanical potential
models \cite{lms_new} and shown to work successfully for the decay
constants~of heavy pseudoscalar mesons \cite{lms_fp}. We introduce
a {\em dual invariant mass\/} $M_{\rm dual}$ and a {\em dual decay
constant\/} $f_{\rm dual}$ by defining
\begin{eqnarray}
\label{mdual}M_{\rm dual}^2(\tau)\equiv-\frac{d}{d\tau}\log
\Pi_{\rm dual}(\tau, s_{\rm eff}(\tau)),\qquad\label{fdual}f_{\rm
dual}^2(\tau)\equiv M_V^{-2}e^{M_V^2\tau}\Pi_{\rm dual}(\tau,
s_{\rm eff}(\tau)).
\end{eqnarray}
For a properly constructed $\Pi_{\rm dual}(\tau, s_{\rm eff}(\tau))$,
the dual mass coincides with the actual ground-state
mass $M_V$. Therefore, any deviation of the dual mass from $M_V$
is an indication of the contamination of the dual correlator by
excited states.

For any trial function for the effective threshold, we derive a
variational solution by minimizing the difference~between the dual
mass (\ref{mdual}) and the actual (i.e., experimentally measured)
mass in the Borel window. This variational solution provides the
decay constant then via (\ref{fdual}). We consider a set of
polynomial Ans\"atze for the effective threshold,~viz.,
\begin{eqnarray}
\label{zeff} s^{(n)}_{\rm eff}(\tau)=
\sum\limits_{j=0}^{n}s_j^{(n)}\tau^{j},
\end{eqnarray}
and fix the coefficients $s_j^{(n)}$ (the knowledge of which then
allows us to compute the decay constant $f_V$) by minimizing
\begin{eqnarray}
\label{chisq} \chi^2 \equiv \frac{1}{N} \sum_{i = 1}^{N} \left[
M^2_{\rm dual}(\tau_i) - M_V^2 \right]^2
\end{eqnarray}
over the Borel window. Still, different Ans\"atze for $s_{\rm
eff}(\tau)$ yield different sum-rule predictions for the decay
constants.

Careful studies of quantum-mechanical potential models indicate
that it suffices to allow for polynomials up to~third order: In
this case, the band delimited by the predictions arising from
linear, quadratic, and cubic Ans\"atze for $s_{\rm eff}(\tau)$
encompasses the true value of the decay constant. Even a good
knowledge of the truncated OPE does not allow~us~to determine the
decay constant precisely but it enables us to provide a range of
values containing the true value of~this decay constant. The width
of this range may then be regarded as the {\em systematic error\/}
related to the principally~limited accuracy of QCD sum rules.
Presently, we are not aware of any other possibility to acquire a
more reliable estimate~for the systematic error. Noteworthy,
considering a $\tau$-independent threshold would not allow us to
probe the accuracy of the obtained estimate for $f_V$.

On top of the systematic error comes the {\em OPE-related error\/}
of the decay constant: the OPE parameters are known only with some
errors, which induce a corresponding error of $f_V$. We determine
this OPE-related (statistical) error by averaging the results for
the decay constant assuming for the OPE parameters Gaussian
distributions with the central values and standard deviations
quoted in (\ref{Table:1}) and a flat distribution over the scale
$\mu$ in the range $3<\mu\;({\rm GeV})<5$.

\subsection{\boldmath Decay constant of the $B^*$ meson}
\subsubsection{Choice of renormalization scale}
In principle, the decay constant should be independent of the
scale $\mu$ at which the correlation function is evaluated. In
practice, however, due to the truncations of the perturbative
expansion and the series of power corrections, and~the neccessity
to isolate the ground-state contribution from the hadron continuum
states, a reliable extraction of the~decay constant may be
performed in only a limited range of the scale $\mu$. For the
vector beauty meson, the suitable range of $\mu$ is found to be
$\mu=3$--5 GeV: For $\mu\le3$ GeV, the perturbative expansion for
the vector correlator does not exhibit a satisfactory perturbative
convergence and therefore gives no reason to believe that the
unknown higher-order radiative corrections both in the
perturbative part of the correlation function and in the radiative
corrections to the condensates are negligible. At higher scales
$\mu\ge5$ GeV, the $B^*$ mass cannot be reproduced with the
required accuracy, signalling that there the contamination of the
excited states cannot be cleaned out.

\subsubsection{Choice of Borel-parameter window}
We require that the $B^*$--$B$ mass splitting and the masses of
$B^*$ and $B$ mesons are reproduced, separately, with an accuracy
not worse than 5 MeV for any $\tau$ value within the selected
ranges. As follows from the properties of the~dual correlators,
this requirement provides two constraints on the choice of the
$\tau$ window for $B^*$:\begin{enumerate}\item The $\tau$ window
for $B^*$ should be chosen at lower values of $\tau$ compared to
the $B$-meson case.\item The precise choice of the $\tau$-window
for $B^*$ should correlate with the scale $\mu$ at which the
correlator is evaluated.\end{enumerate}To satisfy the above
criteria for $B^*$, we set the lower boundary at $\tau_{\rm
min}\;({\rm GeV}^{-2})=0.01$ and choose a $\mu$-dependent~upper
boundary of the form $\tau_{\rm max}\;({\rm GeV}^{-2})=0.31-0.05\,
\mu\;({\rm GeV})$, which choice enables us to extract $f_{B^*}$
with a systematic uncertainty not worse than 5 MeV and strongly
diminishes the unphysical scale dependence of the decay
constant~$f_{B^*}$.

\begin{figure}[ht]
\begin{tabular}{c}
\includegraphics[width=7.5cm]{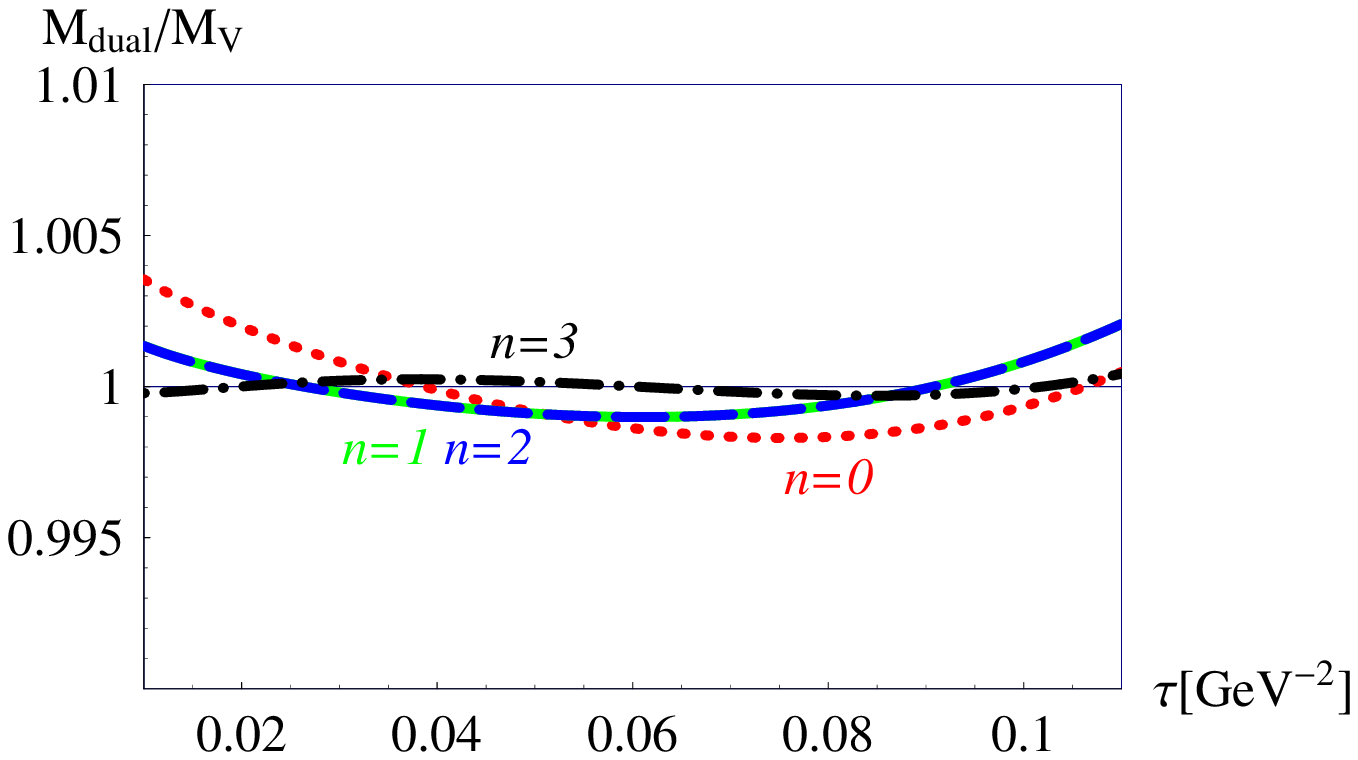} 
\includegraphics[width=7.5cm]{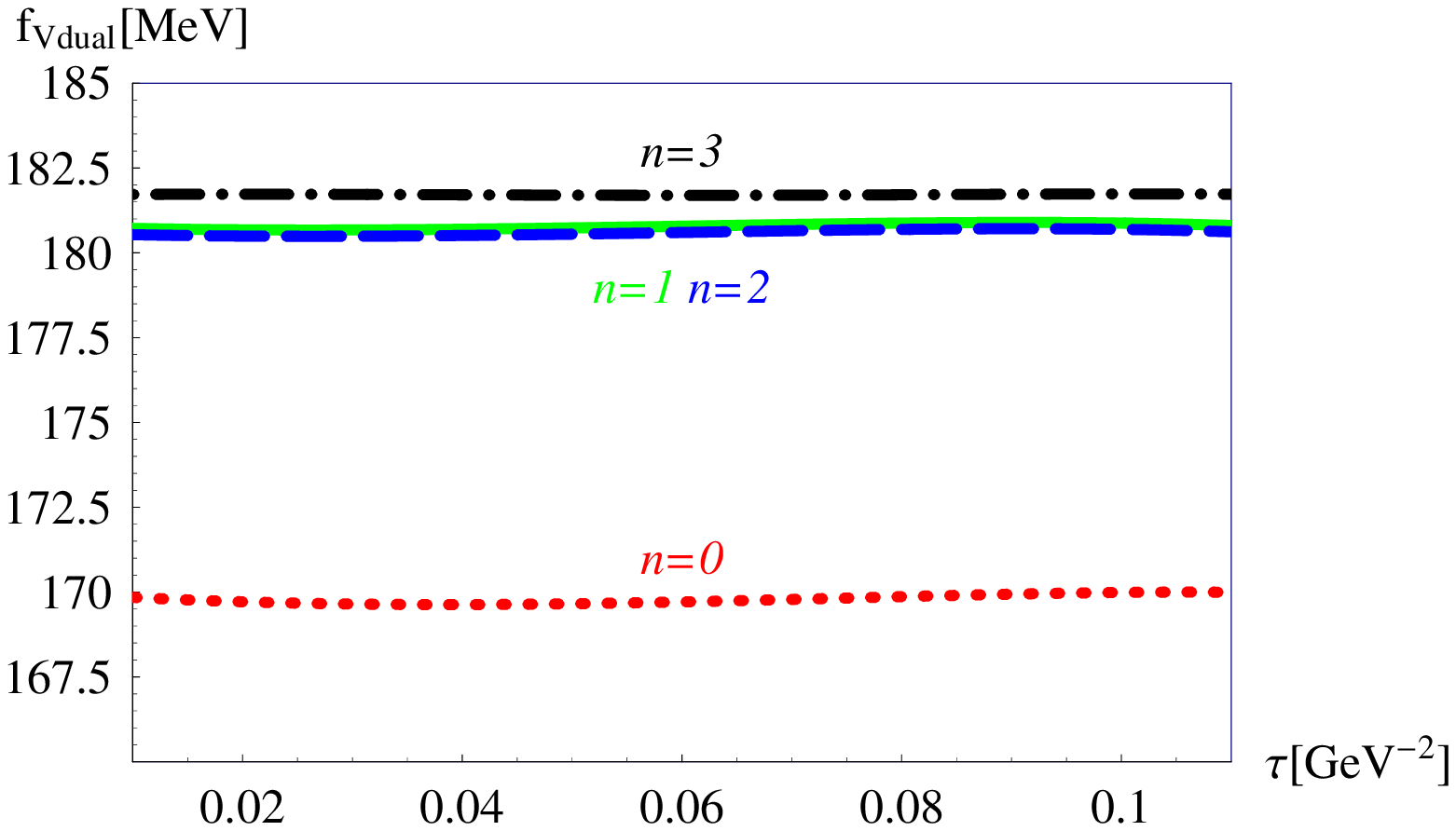}\\
(a) \hspace{7cm} (b)\\
\includegraphics[width=7.5cm]{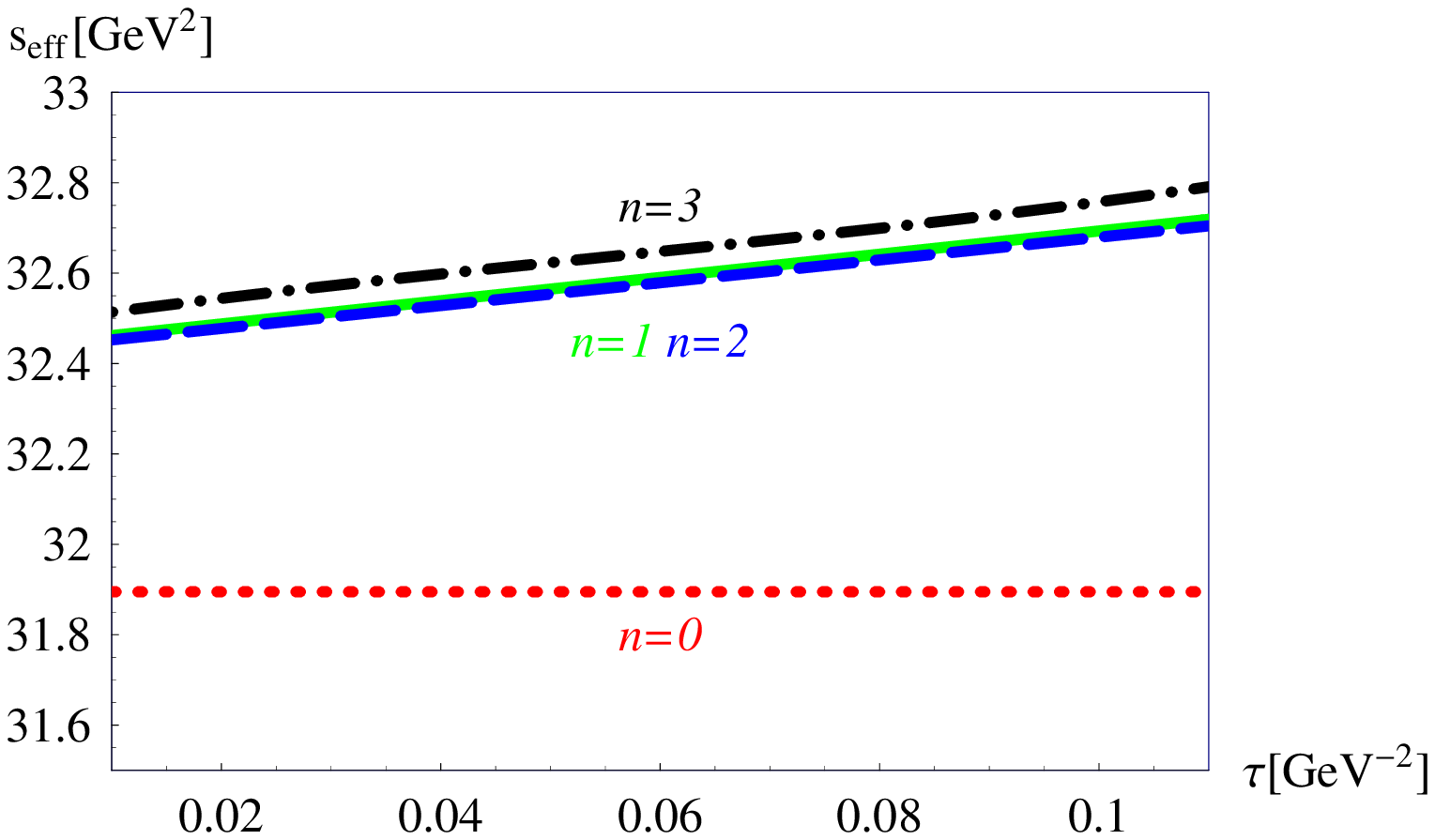}\\
(c)
\end{tabular}
\caption{\label{Plot:fBv} Dependence on the Borel parameter $\tau$
of the dual mass (a) and the dual decay constant (b) of the $B^*$
meson, obtained~by adopting different Ans\"atze (\ref{zeff}) for
the effective threshold $s_{\rm eff}(\tau)$ and fixing these
thresholds by minimizing (\ref{chisq}); the results are presented
for the central values of all OPE parameters. (c) The
$\tau$-dependent effective thresholds as obtained by our
algorithm. The integer $n=0,1,2,3$ is the degree of the polynomial
in our Ansatz (\ref{zeff}) for $s_{\rm eff}(\tau)$: dotted lines
(red)---$n=0$; solid lines (green)---$n=1$; dashed lines
(blue)---$n=2$; dot-dashed lines (black)---$n=3$.}
\end{figure}

Figure~\ref{Plot:fBv} shows the application of our procedure for
fixing the effective threshold and extracting the resulting
$f_{B^*}$. The dependence of our QCD sum-rule result on the
relevant OPE parameters, i.e., the $b$-quark mass
$m_b\equiv\overline{m}_b(\overline{m}_b)$, the quark condensate
$\langle\bar qq\rangle\equiv\langle\bar qq(2\;{\rm GeV})\rangle$
and the gluon condensate $\langle aGG\rangle$, proves to be well
described by a linear~relation:
\begin{align}
\label{fB*}&f_{B^*}^{\rm dual}(m_b,\langle \bar qq\rangle,\langle
aGG\rangle)=(181.8\pm4_{\rm syst})
\left(1-\frac{11}{181.8}\delta_{m_b}\right)
\left(1+\frac{7}{181.8}\delta_{\langle qq\rangle}\right)
\left(1-\frac{1}{181.8}\delta_{\langle aGG\rangle}\right)
\mbox{MeV},
\end{align}
with
\begin{align}
\label{deltas}
&\delta_{m_b}=\frac{m_b-\mbox{4.247\;GeV}}{\mbox{0.034\;GeV}},\qquad
\delta_{\langle qq\rangle}=\frac{|\langle\bar
qq\rangle|^{1/3}-\mbox{0.267\;GeV}}{\mbox{0.017\;GeV}},\qquad
\delta_{\langle aGG\rangle}=\frac{\langle aGG\rangle-0.024\;{\rm GeV}^4}{0.012\;{\rm GeV}^4}.
\end{align}
The above parameters $\delta$ take values between $-1$ and $+1$
when the corresponding OPE quantity varies in its
$1\sigma$~interval. Varying all other OPE parameters in their
$1\sigma$ ranges leads to an effect on $f_{B^*}$ of less than 1
MeV and is not shown~here. Trusting in our experience from exactly
solvable examples, we assume that the systematic uncertainty
interval contains the true value of the decay constant and that
inside this interval the true decay-constant value has a flat
distribution.

As evident from Fig.~\ref{Plot:fBv}(a), using a constant threshold
leads to a contamination of the dual correlator by excited~states
(beyond the acceptable level), while this contamination is
strongly reduced for $n>0$: the values of the decay constant in
Fig.~\ref{Plot:fBv}(b) resulting for $n>0$ are nicely grouped
together, whereas the $n=0$ prediction emerges some 10 MeV~below.

\begin{figure}[ht]
\begin{tabular}{cc}
\includegraphics[width=8.2cm]{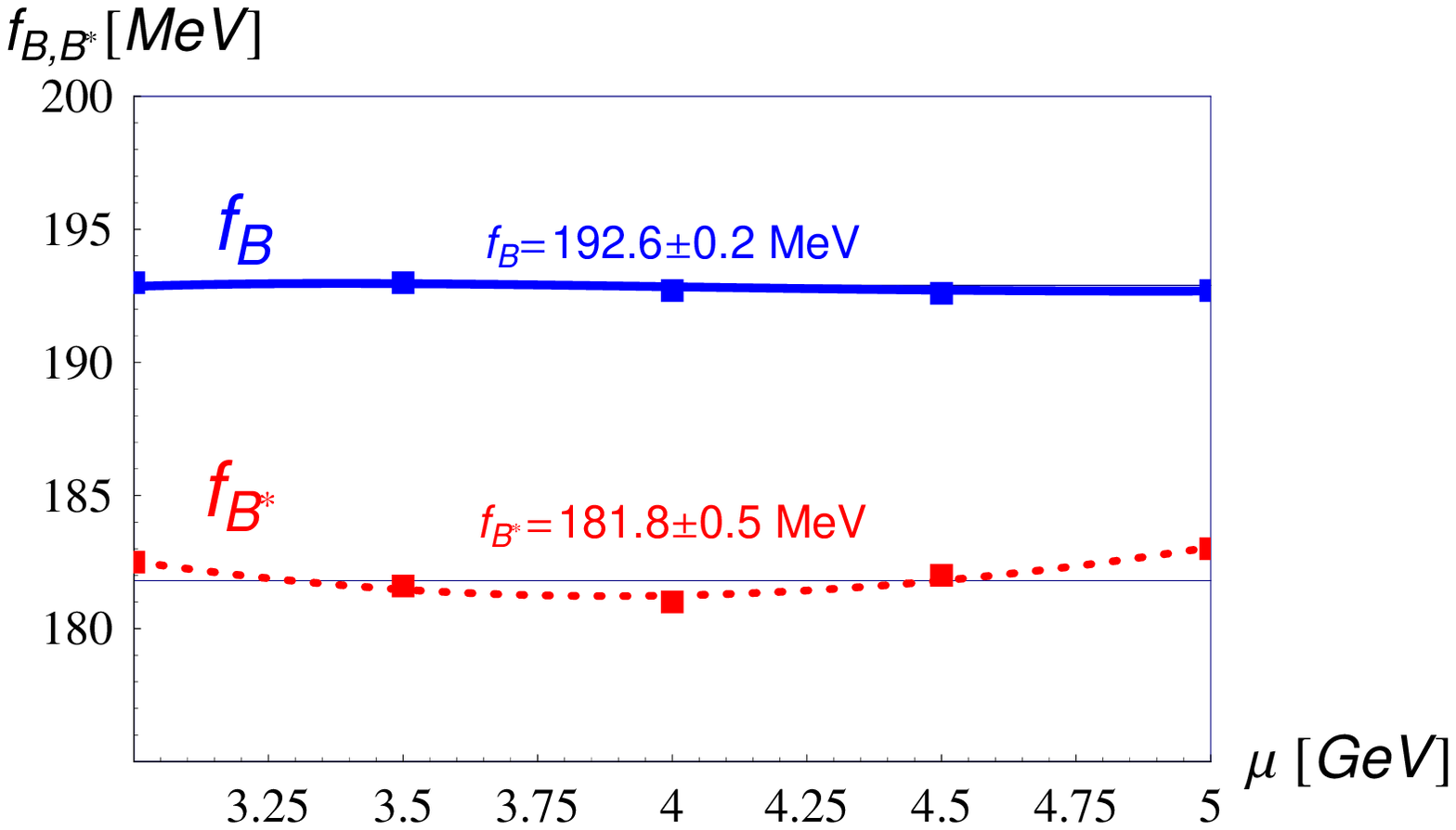}&
\includegraphics[width=8.2cm]{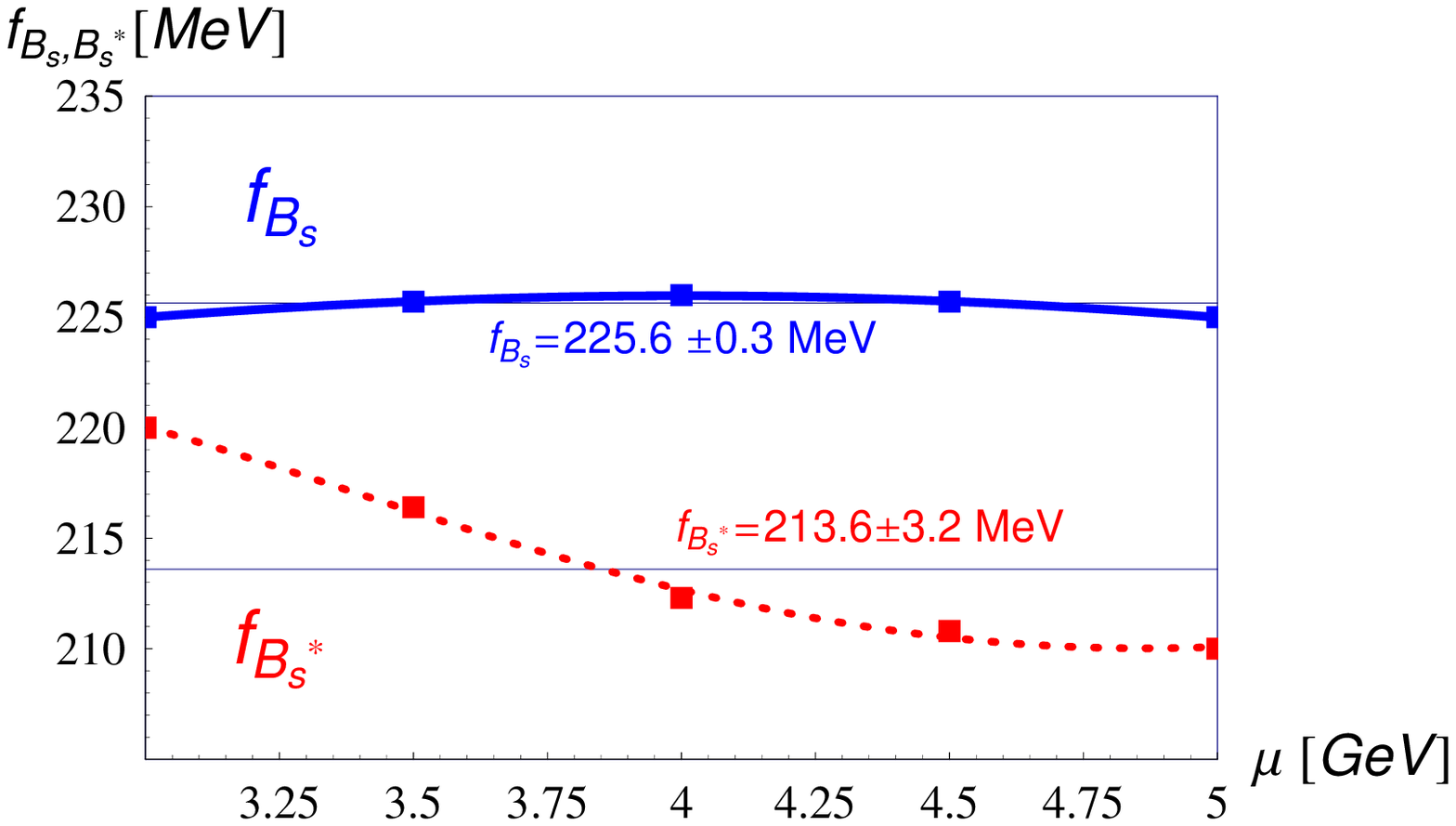}\\(a)&(b)
\end{tabular}
\caption{\label{Plot:fV_vs_mu}
Renormalization-scale dependence of
the predicted decay constants: (a) $f^{\rm dual}_B(\mu)$ and
$f^{\rm dual}_{B^*}(\mu)$, (b) $f^{\rm dual}_{B_s}(\mu)$ and
$f^{\rm dual}_{B^*_s}(\mu)$. For each decay constant, we depict
the $\mu$-related uncertainty, i.e., the standard deviation
calculated assuming a flat $\mu$ distribution in the range
$\mu=3$--5 GeV. Dotted lines (red)---vector beauty mesons; solid
lines (blue)---pseudoscalar beauty mesons.}
\end{figure}

A particularly convincing feature of the presented extraction
procedure is the insensitivity of the extracted value~of $f_{B^*}$
(as well as that of $f_B$) to scale variations in the interval
$\mu=3$--5 GeV (Fig.~\ref{Plot:fV_vs_mu}), achieved by demanding
an accurate reproduction of the $B^*$ mass in the full $\tau$
window, which requires a specific choice of the $\tau$ window
correlated with the scale $\mu$ at which the correlator is
evaluated. Such choice of the $\tau$ window allows us to keep the
systematic uncertainty, estimated by the half width of the band
encompassing the results for the linear, quadratic, and cubic
thresholds, at a level below 4 MeV in the full $\mu$ range.
Therefore, (\ref{fB*}) describes well the result for any $\mu$
from the range $\mu=3$--5~GeV.

\begin{figure}[ht]\begin{tabular}{ccc}
\includegraphics[width=8.5cm]{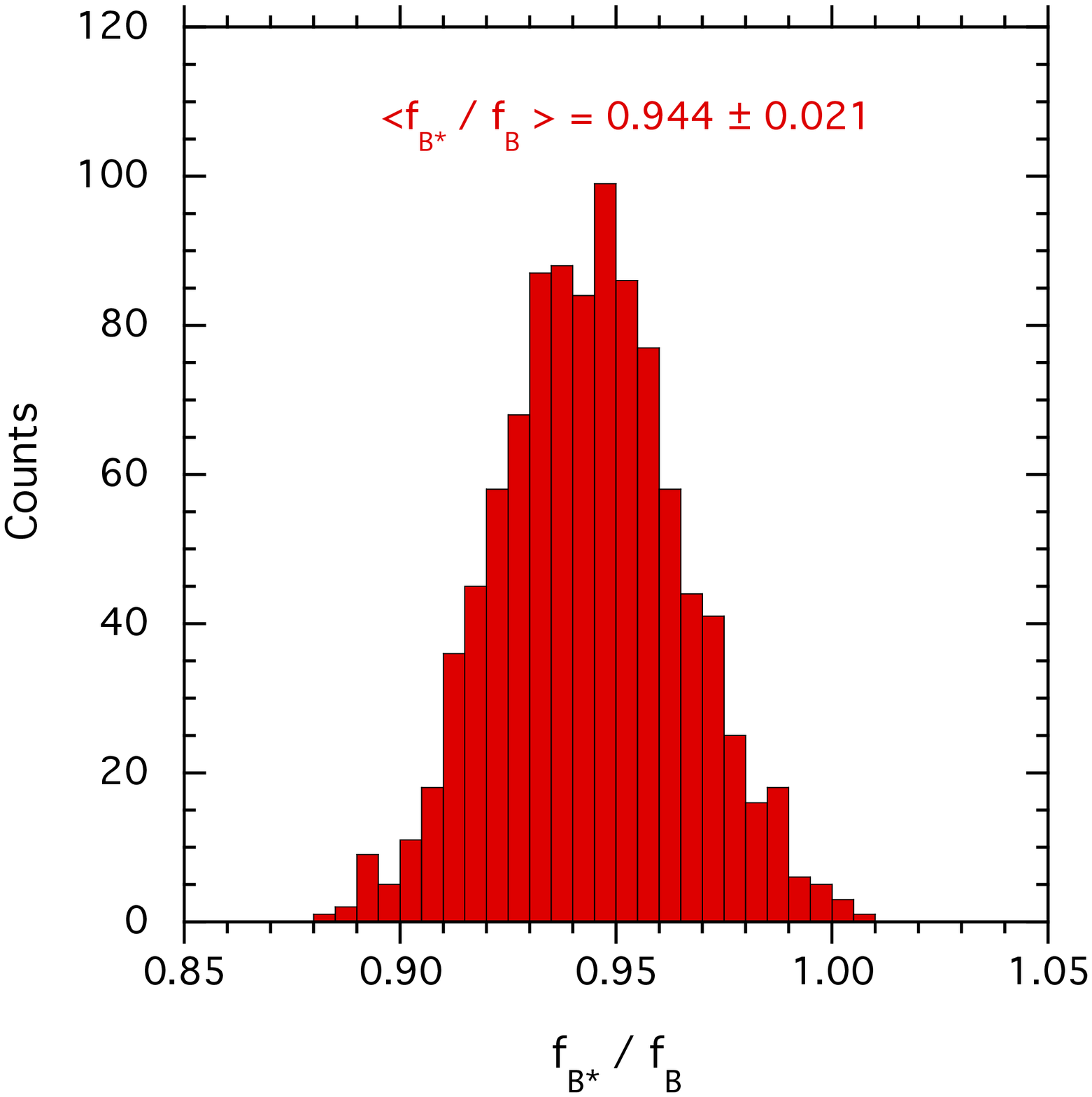}&
\includegraphics[width=8.5cm]{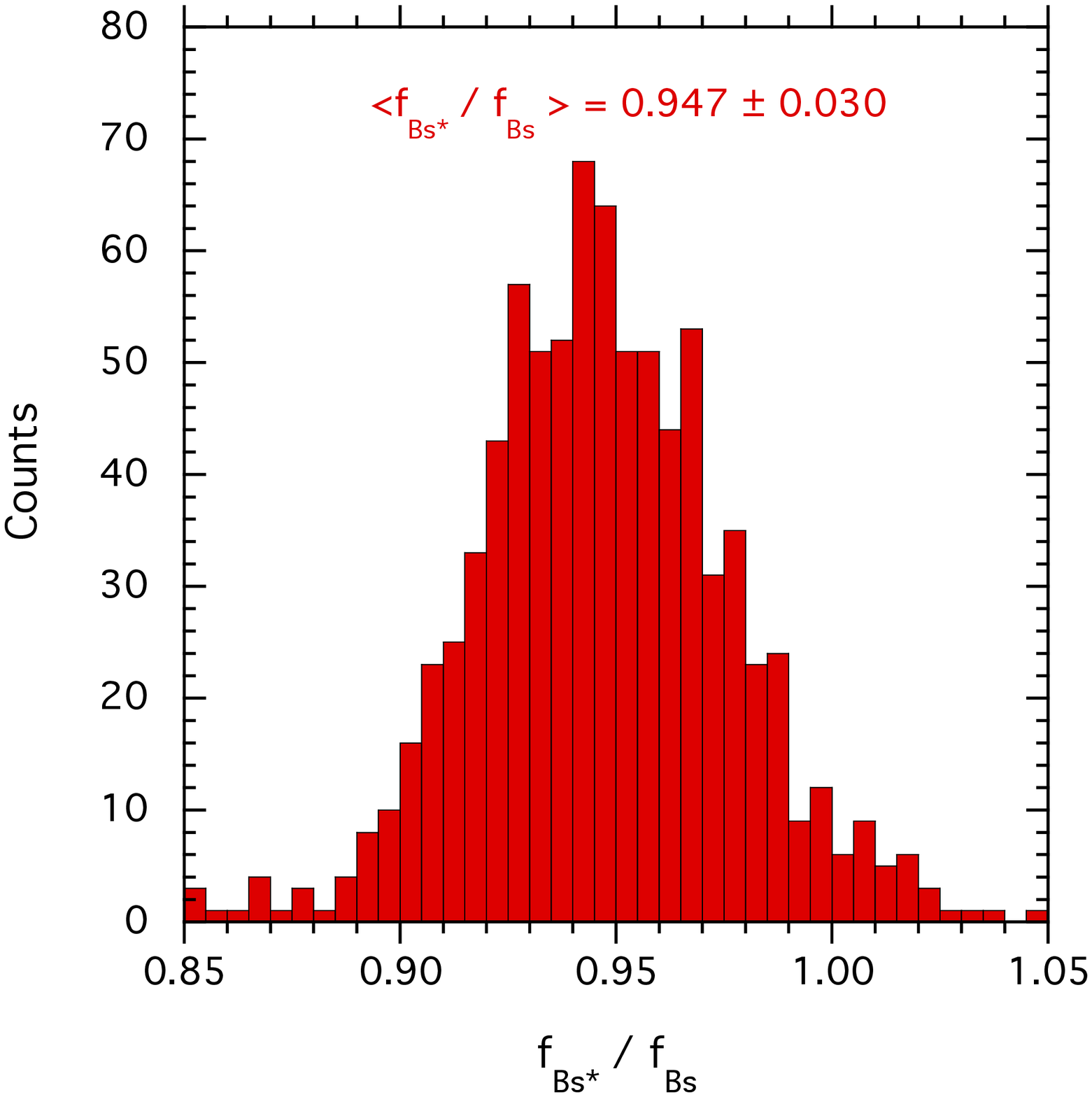}
\end{tabular}
\caption{\label{Plot:bootstrap}Distributions of the ratios
$f_{B^*}/f_B$ and $f_{B_s^*}/f_{B_s}$ of beauty-meson decay
constants, obtained by generating 1000 bootstrap events. For both
ratios, their final distributions possess Gaussian-like shapes,
with the standard deviations quoted in the plots.}
\end{figure}

Assuming Gaussian distributions for all OPE parameters collected
in (\ref{Table:1}), we get the distribution of $f_{B^*}$
depicted~in Fig.~\ref{Plot:bootstrap}. For the average and the
standard deviation of the $B^*$-meson decay constant, we obtain
\begin{eqnarray}
\label{fDv_constant} f_{B^*}=\left(181.8\pm13.1_{\rm OPE}\pm4_{\rm
syst}\right)\mbox{MeV}.
\end{eqnarray}
The OPE uncertainty is composed as follows: 11 MeV are due to the
variation of $m_b$ and 6 MeV arise from the quark condensate. The
uncertainties of all other OPE parameters contribute less than 1
MeV to the OPE uncertainty~of~$f_{B*}$.
The corresponding QCD sum-rule outcome for the $B$-meson decay
constant $f_B$ from our earlier investigation \cite{ourmb} reads
\begin{eqnarray}
\label{fB}f_B^{\rm dual}(m_b,\langle\bar qq\rangle,\langle
aGG\rangle)=(192.6\pm3_{\rm syst})
\left(1-\frac{12.6}{192.6}\delta_{m_b}\right)
\left(1+\frac{6.8}{192.6}\delta_{\langle qq\rangle}\right)
\left(1+\frac{1}{192.6}\delta_{\langle aGG\rangle}\right)
\mbox{MeV}.
\end{eqnarray}
As is obvious from (\ref{fB*}) and (\ref{fB}), the OPE
uncertainties cancel out, to a great extent, in the ratio, which,
consequently, can be predicted with a rather high accuracy:
\begin{eqnarray}
\label{fBvs/fBs}f_{B^*}/f_B=0.944\pm0.011_{\rm OPE}\pm0.018_{\rm syst}.
\end{eqnarray}
The main contribution to the OPE error in the ratio arises from
the gluon condensate, which enters with different~sign in the
pseudoscalar and the vector correlator (in detail:
$\pm0.01_{\langle aGG\rangle}\pm0.005_{m_b}\pm0.001_{\langle
qq\rangle}$). The total uncertainty~of~the ratio is dominated by
the \emph{systematic\/} uncertainties of the decay constants.
Figure~\ref{Plot:bootstrap} shows the distribution of the ratio as
obtained by a bootstrap analysis.

\subsection{\boldmath Decay constant of the $B^*_s$ meson}
For $B^*_s$, we choose the same Borel-parameter window as for
$B^*$ and again require that the deviation of the dual~mass from
the known $B_s^*$ mass does not exceed 10 MeV in the full $\tau$
window. Our findings for the $B^*_s$-meson decay constant may be
cast in the form
\begin{align}
\label{fbsstar}
&f_{B^*_s}^{\rm dual}(\mu=\overline{\mu},m_b,\langle\bar
ss\rangle,\langle aGG\rangle)=(213.6\pm6)
\left(1-\frac{13.2}{213.6}\delta_{m_b}\right)
\left(1+\frac{11.8}{213.6}\delta_{\langle ss\rangle}\right)
\left(1-\frac{1}{213.6}\delta_{\langle aGG\rangle}\right)
\mbox{MeV},
\end{align}
where $\overline{\mu}$ is defined in (\ref{fbsstarmu}) and 

\vspace{-0.5cm}

\begin{eqnarray}
\delta_{\langle ss\rangle}=\frac{|\langle\bar
ss\rangle|^{1/3}-\mbox{0.248\;GeV}}{\mbox{0.033\;GeV}}.
\end{eqnarray}
Unfortunately, the sensitivity of $f_{B_s^*}$ to the choice of the
scale $\mu$ at which the vector correlator is evaluated turns out
to be rather pronounced. This dependence on the choice of $\mu$
may be parametrized by a series in powers of
$\log(\mu/\overline{\mu})$:
\begin{equation}
\label{fbsstarmu}
f_{B_s^*}^{\rm dual}(\mu)=213.6\;\mbox{MeV}
\left[1-0.12\log(\mu/\overline{\mu})+0.11\log^2(\mu/\overline{\mu})
+0.43\log^3(\mu/\overline{\mu})\right],
\qquad\overline{\mu}=3.86\;\mbox{GeV}.
\end{equation}
Averaging over the OPE parameters (using Gaussian distributions of
all OPE parameters except for $\mu$, for which~a~flat distribution
in the range $\mu=3$--5 GeV is assumed) yields
\begin{eqnarray}
\label{fbsstarfinal} f_{B_s^*}=(213.6\pm18.2_{\rm OPE}\pm6_{\rm
syst})\;\mbox{MeV},
\end{eqnarray}
with the following main contributions to the OPE error: 11.5 MeV
from the $s$-quark condensate and 14.1 MeV~from~$m_b$; an
uncertainty of 3.2 MeV arises from the $\mu$ dependence of
$f_{B^*_s}$.

For the pseudoscalar $B_s$ meson, our corresponding estimates read
\begin{eqnarray}
\label{fbs}
f_{B_s}^{\rm dual}(m_b,\langle\bar ss\rangle,\langle aGG\rangle)=
(225.6\pm3_{\rm syst})
\left(1-\frac{14.1}{225.6}\delta_{m_b}\right)
\left(1+\frac{11.5}{225.6}\delta_{\langle ss\rangle}\right)
\left(1+\frac{1}{225.6}\delta_{\langle aGG\rangle}\right)
\mbox{MeV}.
\end{eqnarray}
As seen in Fig.~\ref{Plot:fV_vs_mu}, the sensitivity of $f_{B_s}$
to the choice of $\mu$ is negligible. The total OPE uncertainty of
$f_{B_s}$ is rather~large:
\begin{eqnarray}
\label{fbsfinal}f_{B_s}=(225.6\pm18.3_{\rm OPE}\pm3_{\rm
syst})\;\mbox{MeV}.
\end{eqnarray}
The decomposition of the OPE error reads: 11.5 MeV are due to the
error of $s$-quark condensate and 14.1 MeV due to the error of
$\overline{m}_b(\overline{m}_b)$, the uncertainties of the other
OPE parameters contribute at the level of 1 MeV.

Similar to the $f_{B^*}/f_B$ case, except for the gluon-condensate
contribution the OPE uncertainties cancel, to a great extent,~in
the ratio of the decay constants, which may thus be predicted
rather accurately:
\begin{eqnarray}
\label{bsstarbsfinal} f_{B_s^*}/f_{B_s}=0.947\pm0.023_{\rm
OPE}\pm0.020_{\rm syst}.
\end{eqnarray}
The OPE uncertainty in the ratio is dominated by the sensitivity
of $f_{B_s^*}$ to the choice of the scale $\mu$. The (obligatory)
bootstrap analysis gives for the ratio $f_{B_s^*}/f_{B_s}$ the
nearly Gaussian distribution shown in Fig.~\ref{Plot:bootstrap}.

\section{Summary and conclusions}
Exploiting the tools offered by QCD sum rules, we analyzed in
great detail the decay constants of the beauty vector mesons,
paying special attention to the uncertainties arising in our
predictions for the decay constants: the OPE~error, related to the
precision with which the QCD parameters are known, and the
systematic error, intrinsic to the QCD sum-rule approach as a
whole, reflecting the limited accuracy of the extraction
procedure. Our findings are as follows:

\begin{itemize}
\item[(i)]
As was already noted in the case of heavy pseudoscalar mesons
\cite{lms_fD}, also for the vector correlator the perturbative
expansion in terms of the heavy-quark pole-mass does not exhibit
good convergence. Reorganizing the OPE in terms of the
corresponding running mass allows us to choose a range of scales
for which, upon evaluation of the correlator, the perturbative
hierarchy becomes explicit. For scales $\mu\le2.5$--3 GeV, also
the running-mass OPE does not exhibit any hierarchy of
perturbative contributions; at too large scales $\mu\gtrsim5$--6
GeV, we observe a strong cancellation between the large positive
zero-order and the large negative first-order contributions,~thus
signalling that the accuracy of the OPE may deteriorate. There is,
however, a sizeable interval of scales, $3\le\mu\;({\rm
GeV})\le5$, where the $O(a^2)$-truncated OPE provides a good
description of the dual correlation function.

\item[(ii)]
Requiring the known value of the meson mass to be well reproduced
in a relatively broad $\tau$ window leads, in~the case of the
vector mesons, to some correlation between the scale $\mu$ at
which the correlator is evaluated and the upper boundary of the
$\tau$ window: for $\mu\gtrsim5$ GeV, the Borel window for the
vector correlator shrinks and thus no meaningful extraction of the
decay constants of $B^*$ and $B_s^*$ from sum rules is possible.
The observed correlation between the parameters of the Borel
window and the value of $\mu$ strongly reduces the (unphysical)
$\mu$ dependence of the extracted beauty-meson decay constants.

\item[(iii)]
The $\tau$-dependence of the effective threshold and the details
of the algorithm for fixing this quantity are crucial for
obtaining realistic estimates of the systematic uncertainty of the
extracted decay constant. For the analysis~of the ratios of the
decay constants of vector to pseudoscalar beauty mesons, where the
mass splitting between~the vector and the pseudoscalar partners
amounts to some $45$ MeV only, the stringent requirement to
reproduce~this splitting and the individual masses of vector and
pseudoscalar beauty mesons with an accuracy not worse than 5 MeV
in the full $\tau$ range is crucial for obtaining the low
systematic uncertainty of the extracted decay constants.

\item[(iv)]
The decay constants of pseudoscalar and vector beauty mesons
exhibit a strong dependence on the precise value of
$\overline{m}_b(\overline{m}_b)$. Therefore, the $B_{(s)}$ and
$B^*_{(s)}$ decay constants suffer from large OPE uncertainties.
The~systematic uncertainties of the extracted decay constants are
of the level of a few MeV and remain under good control.

\item[(v)]
The ratios $f_{B^*}/f_B$ and $f_{B_s^*}/f_{B_s}$ can be predicted
with very good accuracy because of large cancellations between the
OPE uncertainties in the ratios and a good control over the
systematic uncertainties of the decay constants. Our final results
read
\begin{eqnarray*}
f_{B^*}/f_{B}=0.944\pm0.021,\qquad
f_{B_s^*}/f_{B_s}=0.947\pm0.030,
\end{eqnarray*}
where the error given is the total uncertainty, including the
systematic and the OPE uncertainty. The resulting distributions
are close to normal distributions (Fig.~\ref{Plot:bootstrap}),
thus the quoted errors are Gaussian standard deviations.

\item[(vi)]
Our results are in excellent agreement with and have a precision
comparable to the recent lattice QCD values~\cite{hpqcd2015}
\begin{eqnarray*}
f_{B^*}/f_{B}=0.941\pm0.026,\qquad
f_{B_s^*}/f_{B_s}=0.953\pm0.023.
\end{eqnarray*}
\end{itemize}

\vspace{3ex}\noindent{\bf Acknowledgements.} The authors thank
B.~Blossier, O.~P\`ene, H.~Sazdjian, and B.~Stech for interesting
discussions. D.~M.~is grateful to the Alexander von
Humboldt-Stiftung (Germany) for supporting his stay in Heidelberg,
where part of this work was done. S.~S.~thanks MIUR (Italy) for
partial support under contract No.~PRIN 2010--2011.

\end{document}